# Optical absorption of excimer laser-induced dichlorine monoxide in silica glass and excitation of singlet oxygen luminescence by energy transfer from chlorine molecules

Linards Skuja*, Nadège Ollier, Koichi Kajihara, Ivita Bite, Madara Leimane, Krisjanis Smits, Andrejs Silins

Dr. L. Skuja, I. Bite, M. Leimane, Dr. K. Smits, Prof. A. Silins
Institute of Solid State Physics, University of Latvia
8 Kengaraga str., LV1063, Riga, Latvia
E-mail: skuja@latnet.lv

Dr. Nadège Ollier
Laboratoire des Solides Irradiés CEA-CNRS-Ecole Polytechnique UMR7642
Institut Polytechnique de Paris, 91128,
Palaiseau Cedex, France.

Prof. Koichi Kajihara
Department of Applied Chemistry for Environment
Graduate School of Urban Environmental Sciences
Tokyo Metropolitan University
1-1 Minami-Osawa, Hachioji, Tokyo 192-0397, Japan

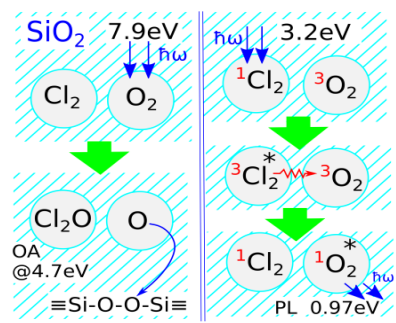

**Abstract**

An optical absorption (OA) band of interstitial dichlorine monoxide molecules with peak at 4.7 eV and halfwidth 0.94 eV is identified in $F_2$ laser – irradiated ($\hbar\omega$=7.9 eV) synthetic silica glass bearing both interstitial $O_2$ and $Cl_2$ molecules. Alongside with intrinsic defects, this OA band can contribute to solarization of silica glasses produced from $SiCl_4$. While the formation of ClClO is confirmed by its Raman signature, its structural isomer ClOCl may also contribute to this induced OA band. Thermal destruction of this band between 300°C and 400°C almost completely restores the pre-irradiation concentration of interstitial $Cl_2$. An additional weak OA band at 3.5 eV is tentatively assigned to $ClO_2$ molecules. The strongly forbidden 1272nm infrared luminescence band of excited singlet $O_2$ molecules was observed at 3 eV-3.5 eV excitation, demonstrating an energy transfer process from photoexcited triplet $Cl_2$ to $O_2$. The energy transfer most likely occurs between $Cl_2$ and $O_2$ interstitial molecules located in neighboring nanosized interstitial voids in the structure of $SiO_2$ glass network.

1. Introduction

Synthetic $SiO_2$ glass is the dominant optical material for ultraviolet (UV) applications. The decrease of its transparency due to UV-irradiation ("solarization") is caused by generation of point defects. The separate components of the typically broad UV optical absorption (OA) spectrum in the 4-7 eV range are still not completely identified. The main UV-absorbing intrinsic defects are oxygen dangling bonds ("non-bridging oxygen hole centers", NBOHCs) with broad OA bands at 4.8 eV and higher energies[1] and different variants of silicon dangling bonds (E'-centers) with OA bands at 5.8 eV±0.1eV [2]. At higher energies, additional absorption is caused by bound hydroxyl (silanol, $\equiv$Si−O−H) or chloride groups ($\equiv$Si−Cl), which are typically present in (often referred as "high-purity") synthetic silica" due to the manufacturing from $SiCl_4$ precursor and optional use of $H_2$-$O_2$ flame pyrolysis. This OA is unproblematic for many UV applications, since its onset is at photon energies in vacuum UV range: $\hbar\omega$ > 7.3 eV for SiOH groups [3–5] and ≥ 6.7 eV for $\equiv$Si−Cl groups [6]. However, in "dry" (low OH concentration) silicas, part of chlorine impurities form interstitial $Cl_2$ molecules, which give rise to a weak near-UV OA band at 3.8 eV [6] and to an associated structured $Cl_2$ photoluminescence (PL) band at 1.23 eV [7].

The peak absorption cross section σ of the 3.8 eV OA band is relatively small ($2.6\times10^{-19}$ $cm^2$ for free $Cl_2$ molecules). Therefore this band is of practical importance mostly in fiber optics geometry, where it severely limits the UV transparency of commercial dry silica multimode optical fiber waveguides for visible-to-infrared applications (e.g., [8]). Another weak chlorine-related OA band appears at 3.26 eV in irradiated optical fibers, which is tentatively assigned [9,10] to radiation-induced interstitial chlorine atoms $Cl^0$, previously identified by EPR.[11]

However, chlorine-related species with possibly much stronger OA can form in silica. In the co-presence of interstitial $Cl_2$ and $O_2$ molecules, creation of dichlorine monoxide ClClO molecules by 157 nm irradiation was detected in Raman spectra. [7] OA band of ClClO isolated in Ar matrix at T=17 K is located at λ=260 nm (4.77 eV) with a large peak absorption cross section σ=$1.3 \times 10^{-17}$ $cm^2$. [12] It is ≈50 times higher than σ of 3.8 eV OA band of $Cl_2$, and therefore ClClO creation in principle can significantly contribute to the UV solarization of silica. While there is no direct Raman evidence for the other structural isomer of dichlorine monoxide, ClOCl in silica glass, its formation and possible contribution to solarization should be considered too.

The purpose of this work was to elucidate the UV absorption band of ClClO, ClOCl or other chlorine oxide molecules in silica glass, and to investigate the conditions under which they can form.

## 2. Experimental
### 2.1 Samples and irradiation.

Oxygen-excess "dry" silica glasses, synthesized by oxidizing $SiCl_4$ in oxygen plasma were used. They contained relatively large concentrations of interstitial $O_2$ and $Cl_2$ molecules and traces of silanol groups (see Table 1, Samples "A", "B"..."F"). The concentration estimates of $O_2$ were based on Raman/PL spectra, as described in ref. [13]; an approximate estimate of $Cl_2$ was obtained from the amplitude of the 3.8 eV OA band, using peak absorption cross section σ=$2.58 \times 10^{-19}$ $cm^2$ for $Cl_2$ in gas phase.[14] The possible (≈20%) increase of σ due to effective field in $SiO_2$ matrix was neglected. Samples A-E were experimental ones [15]; sample F was of commercial "Suprasil W1" type. Historically, glass of this type has served for numerous past studies as a "benchmark" type of dry synthetic silica glass. Concentration of $Cl_2$ in this sample was below our detection limit. Samples were optically polished, thickness 5mm (A-E) and 2.54 mm (F).

Sample "A" was irradiated in vacuum for 120 s with 157 nm (7.9 eV) photons using $F_2$ excimer laser (Lambda Physik LPF-210) emitting ≈20ns long pulses with power density 19.6 mJ/cm$^2$ at repetition rate 50 pulses/s. That yielded the cumulative fluence of 11.74 J/cm$^2$ or 9.27×10$^{19}$ photons/cm$^2$. To remove the slight surface contamination caused by photolysis of residual vapors in the vacuum chamber, the sample was etched in diluted (1%) HF for 30 s.

**Table 1.** Concentrations of interstitial $O_2$, $Cl_2$ molecules and bound hydroxyl groups SiOH in synthetic $SiO_2$ glass samples used. Concentrations are given in $10^{17}$ cm$^{-3}$ units. The right-hand column gives relative intensities of singlet $O_2$ infrared PL (1272 nm, 0.974 eV), excited at 385 nm.

| Name | [$O_2$] | [$Cl_2$] | [SiOH] | PL [$O_2$] |
|------|------|-------|--------|---------|
| A    | 4.3  | 10.4  | 0.5    | 41.9    |
| B    | 4.2  | 6.4   | 0.5    | 16.9    |
| C    | 2.9  | 6.6   | 0.5    | 21.15   |
| D    | 4.5  | 1.3   | 0.5    | 4.91    |
| E    | 2.8  | 0.7   | 0.5    | 1.95    |
| F    | 3.2  | < 0.2 | <0.5   | 1.07    |

**2.2 Instrumentation.**

Luminescence excitation sources were 3rd harmonic (355 nm, 3.49 eV) of Q-switched Nd-YAG laser (Spectra Physics Quanta Ray/Indi), pulse length 10 ns, energy in the range 1.5 mJ, 4.4 mJ; 385nm UV LED (Nichia U385) with 800 mW flux, 766 nm diode laser (Leading Tech ADR1805, maximum 600 mW) UV LEDs.

Singlet $O_2$ singlet luminescence band at 1270 nm was measured using 355 nm excitation and Andor SR193 spectrograph with DU490 InGaAs CCD camera (spectral resolution 8 nm). For 385 nm and 766 nm excitation PL spectra and PL decay curves were measured by a scanning 200 mm monochromator (spectral resolution 5 nm) and Hamamatsu-R5509-43 liquid $N_2$-cooled near-infrared photomultiplier tube (PMT) or cooled photodiode (G12180-220A). The slow (seconds-range) PL decay was measured by pulsed excitation and signal integration by multichannel photon counter (Fastcomteh 7822) or by analog waveform recorder (Agilent 34410A). Pulsed excitation was provided by mechanical shutter (766nm) or by driving LED current (385 nm).

Optical absorption spectra were obtained using Hamamatsu 10082CAH CCD spectrograph with spectral resolution 1.5 nm and deuterium lamp (Ocean Optics). All measurements were performed at room temperature.

Raman spectra were taken with an aim to detect small irradiation-induced changes on the background of the fundamental Raman spectrum of silica glass. This required particularly high temporal stability and low signal/noise ratio. The spectra were taken in back-scattering geometry using 532 nm laser and CCD spectrograph (Andor Newton/Shamrock303). Details of the hardware and the measurement procedure are described in ref.[7]

## 3. Results
### 3.1 Optical absorption (OA)

OA spectra of pristine and $F_2$ laser-irradiated sample ("A" in **Table 1**) are shown in **Figure 1**. Non-irradiated sample (trace 1) shows only the well-known Gaussian-shaped OA band of interstitial $Cl_2$ molecules at 3.78 eV with full width at half maximum (fwhm) 0.70 eV [14,16,17] and a shoulder of OA bands deeper in vacuum-UV. $F_2$ laser irradiation gives rise to a distinct band at ≈4.8 eV and to a slight decrease at 3.8 eV (trace 2). In the induced (difference) OA spectrum (trace 3) this decrease is more distinct, causing a negative-going peak at 3.8 eV. It can be compensated for by adding an appropriately scaled (amplitude 0.06 cm$^{-1}$) Gaussian band with peak energy (3.78 eV) and fwhm (0.70 eV), corresponding to OA of $Cl_2$ molecules (trace 4). The low-energy wing of the resulting sum spectrum (trace 5) can be approximated by a Gaussian (trace 6) with peak at 4.72 eV, fwhm 0.83 eV.

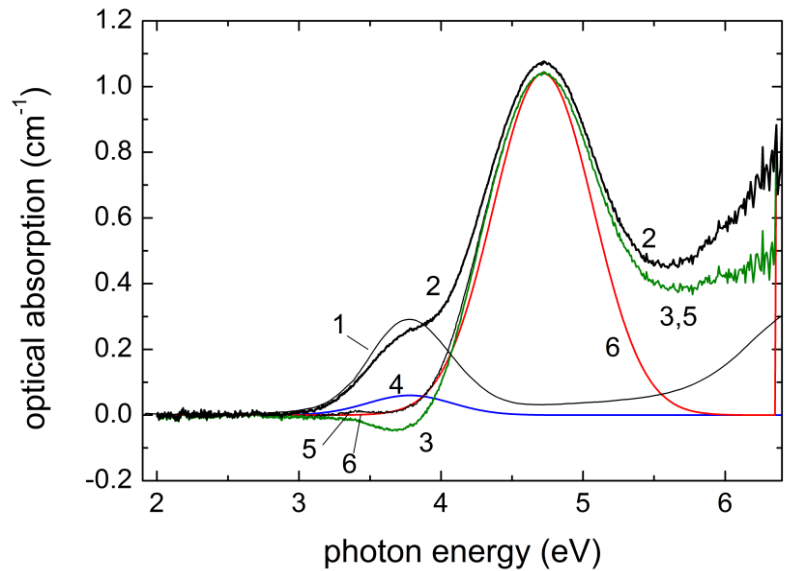

**Figure 1.** Optical absorption spectra of pristine and $F_2$-laser-irradiated synthetic silica glass containing interstitial $O_2$ and $Cl_2$ molecules ("sample A"): (1) - before irradiation, (2) after irradiation, (3) induced absorption (difference) spectrum ("3"="2" − "1"), (4) Gaussian component, with amplitude 0.06 cm$^{-1}$, corresponding to the absorption of $Cl_2$ molecules, destroyed by irradiation, (5) induced absorption, compensated for the destruction of $Cl_2$ ("5"="3"+"4"), (6) Gaussian approximation to the low energy wing of spectrum "5": peak 4.72 eV, fwhm 0.83 eV.

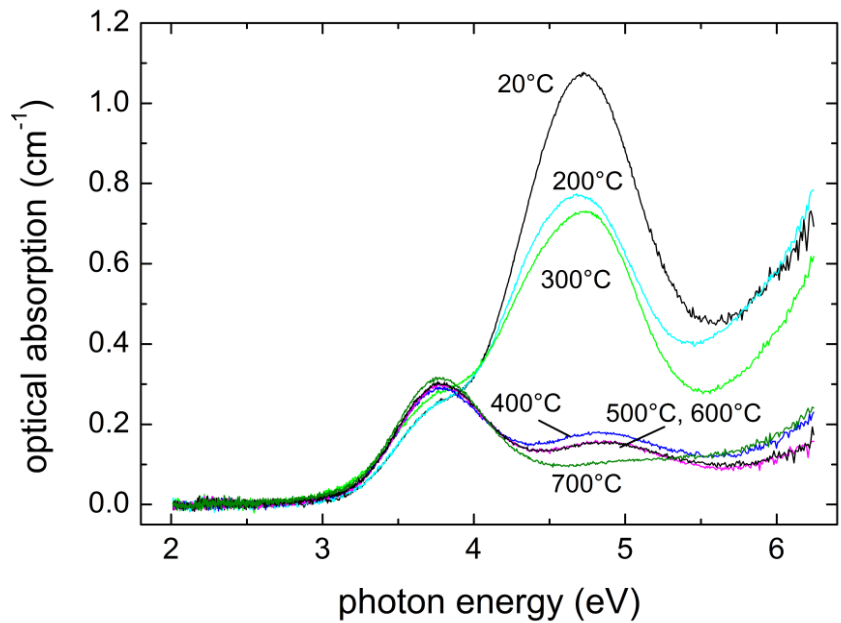

**Figure 2.** Isochronal annealing of optical absorption spectra in $F_2$-laser irradiated chlorine-rich silica glass ("sample A"). The spectra are measured after holding the sample for 15 minutes at temperatures incremented in 100°C steps from 200°C to 700°C. The spectrum taken before the annealing is denoted by "20°C".

The irradiated sample A was thermally annealed by holding for 15 minutes at each annealing temperature, which was increased from 200°C to 700°C in steps of 100°C. OA and Raman spectra (see below) were taken at each step. **Figure 2** shows the changes in OA spectra. It is evident that the strongest changes occur in two temperature intervals (20°C...200°C) and (300°C... 400°C). **Figure 3**, trace 1 shows the OA component, destroyed by annealing at 200°C. It has nearly Gaussian shape (trace 2) with peak at ≈4.8 eV and fwhm 0.8 eV. No significant changes occur in the 3-4 eV spectral range, notably, the 3.8 eV OA band of $Cl_2$ remains unchanged. Annealing at 300°C brings relatively small changes in OA spectrum (**Figure 3**, trace 3), there is a negative dip at ≈ 3.5 eV meaning that some absorbing species are *created* by annealing. **Figure 4**, trace 1 shows the strong OA component, annealed out between 300°C and 400°C. It is nearly Gaussian and, at the first glance, similar to the main OA component annealed at 200°C (**Figure 3**, trace 1). However, it is slightly red-shifted (peak at 4.7 eV) and broader (fwhm 0.94 eV). A more distinctive difference from the OA component, annealed at 200°C is the negative peak at 3.8 eV, which shows that OA band with a peak close to this energy is *restored* by annealing.

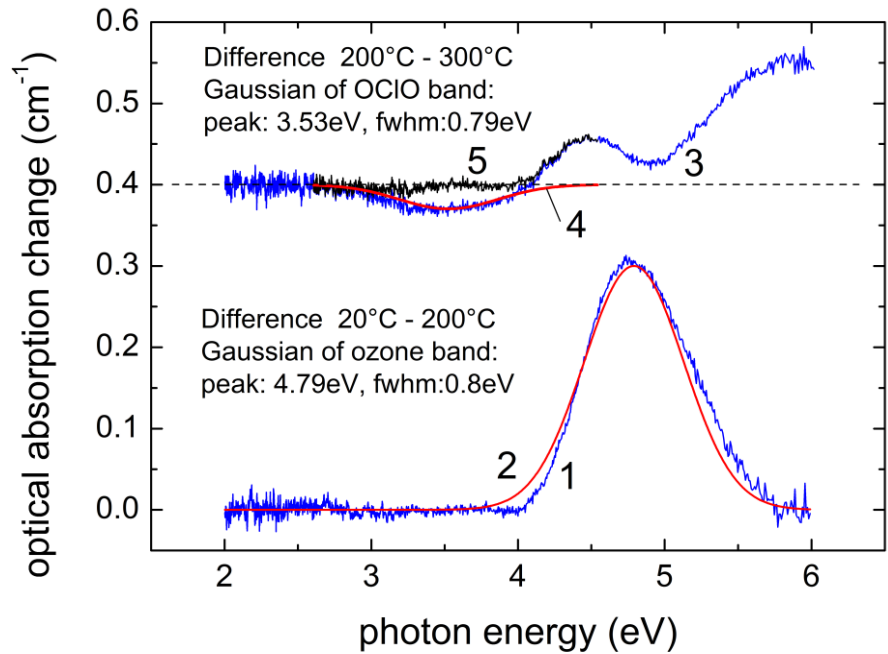

**Figure 3.** Bottom part: Optical absorption component destroyed by thermal annealing at 200°C (trace 1) and its comparison with Gaussian curve drawn with parameters corresponding to the OA band of interstitial ozone $O_3$ molecules. Top part (up-shifted by 0.4 cm$^{-1}$): Changes of optical absorption due to 300°C annealing step (trace 3), negative Gaussian with peak/fwhm corresponding to the OA spectrum of O-Cl-O molecules (trace 4) and with an amplitude scaled to match the negative dip in the measured spectrum (3); difference spectrum (trace 5) = (trace 3- trace 4).

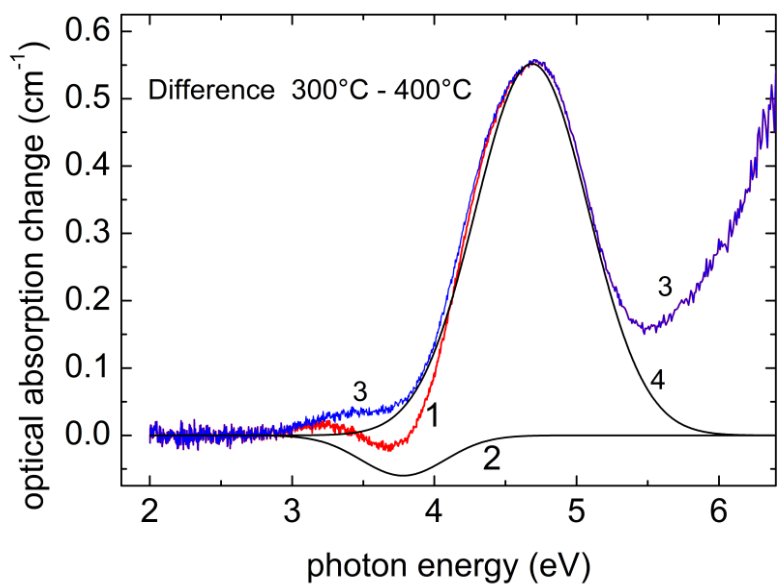

**Figure 4.** Optical absorption in "high-$Cl_2$" sample "A" destroyed by thermal annealing between 300°C and 400°C: Difference spectrum (trace 1), the putative Gaussian component due to the recovery of interstitial $Cl_2$ (trace 2, see text), difference spectrum, compensated for the recovery of $Cl_2$ (trace 3=(trace 1)-(trace 2)), Gaussian fit to the main peak of trace 3 (trace 4 , peak: 4.686 eV, fwhm:0.935 eV, height=0.552 $cm^{-1}$).

**3.2 Raman scattering spectra**

Alongside with OA measurements, Raman spectra of sample "A" were taken after each isochronal thermal annealing step. Since the purpose was to identify the minor signatures of the decaying species, an utmost care was taken to ensure repeatability and low noise (see ref.[7] for details). The bottom trace in **Figure 5** shows a typical Raman spectrum, which, viewed at this magnification scale, remained the same during the entire isochronal annealing series. However, the fine differences between spectra taken before and after each annealing step, plotted at ~50× magnification show a number of bands (**Figure 5**, right side). Some of them, in particular the "peak derivative"-shaped bands are evidently related to slight spectral shifts of the fundamental Raman bands. The largest ones, which occur into the 300 $cm^{-1}$-600 $cm^{-1}$ range, are not shown here. The exact spectral positions of these bands are very sensitive to the accuracy of intensity normalization before performing the subtraction of the spectra. In contrast, the difference spectra can be calculated reliably in the flat spectral regions of low fundamental scattering intensity. Two narrow lines can be identified in the difference spectra: at 1550 cm$^{-1}$ and at 955 cm$^{-1}$ (**Figure 5**). Their intensities are very weak. For example, the integral intensity of the 955 cm$^{-1}$ line (marked by red circle in **Figure 5**) is ≈2×10$^{-5}$ from the integral intensity of the entire Raman spectrum.

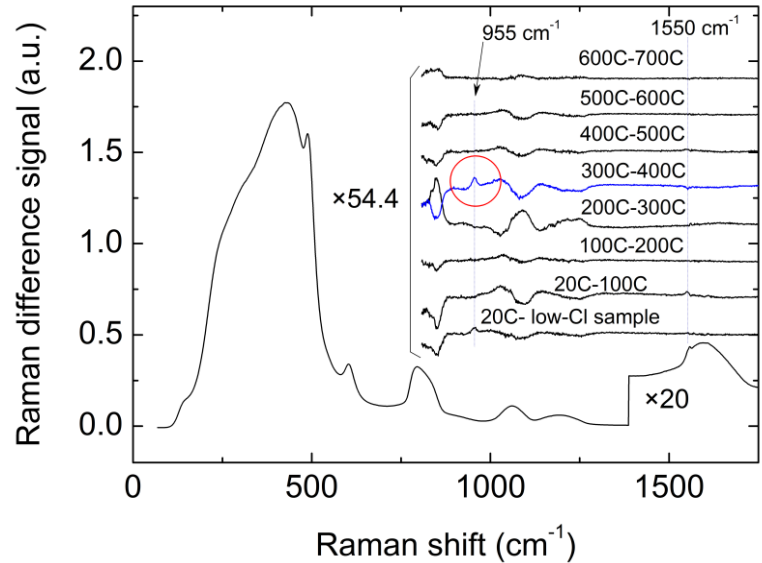

**Figure 5**. Effects of 15 minute isochronal annealing steps on Raman spectrum of F$_2$-laser irradiated silica glass ("sample A"). The bottom trace represents the fundamental Raman spectrum, which remains almost unchanged by the annealing. The subtle differences introduced by the annealing are shown magnified (×54.4) above at right. The lowest difference spectrum ("20C – low-Cl") is obtained by subtracting the Raman spectrum of a similarly sized sample with a lower Cl$_2$ concentration (sample "D" in **Table 1**).

### 3.3 Infrared photoluminescence (PL)

During the Raman experiments with samples A-F it was unexpectedly observed that under 3rd harmonic pulsed Nd-YAG laser excitation (355 nm, 3.49 eV) they emit infrared PL at 1272 nm (0.97 eV) (**Figure 6**, trace 1). This PL band is a fingerprint of widely studied species, an excited metastable singlet state O$_2$ molecule (see, e.g., paper [18] for overview and references). It is also much studied in silica (e.g., paper [19]). Its excitation in 300 nm-400 nm spectral region was unexpected, since O$_2$ molecule has no electronic states there.[20] By varying the pulse intensities it was verified that the dependence of PL intensity on pulse energy is slightly sub-linear, ruling out the 2-photon excitation. The signal PL was stable in time, photobleaching or enhancement of signal under laser irradiation was not observed. Further, this PL band was observed also with 385 nm (3.22 eV) CW LED excitation (trace 3)

and – as expected- with the previously reported (e.g., [18,19]) "direct" 766 nm (1.618 eV) laser excitation of $O_2$ (trace 2).The slight differences in the PL band fwhm and peak positions evident in **Figure 6** are not meaningful at the present accuracy level. The larger measured halfwidth (fwhm) of PL excited at 355 nm (trace 1) is probably due to the lower spectral resolution in this case.

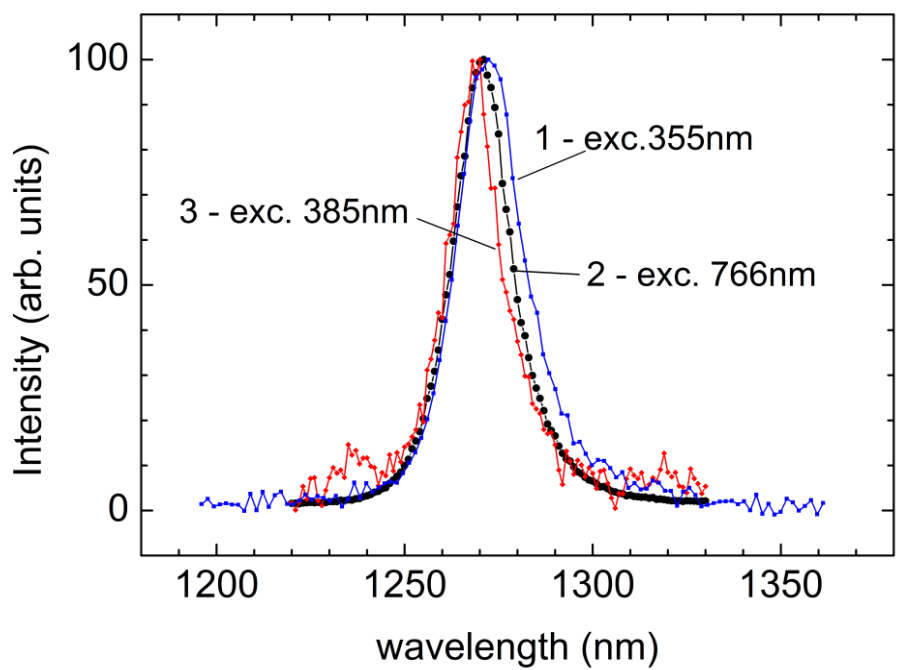

**Figure 6.** Infrared photoluminescence spectra of sample B, measured at room temperature with excitation by: (1) pulsed laser at 355 nm, (2) CW laser at 766 nm, (3) CW LED at 385 nm. Spectral resolution: 8 nm, 5 nm, 5 nm, respectively.

The intensities of the 1272 nm PL band in different samples (A-F) were checked using the 385 nm excitation, and the results were plotted in correlation with the respective concentrations of $Cl_2$ and $O_2$ in the samples (**Figure 7**, panels 1 and 2).

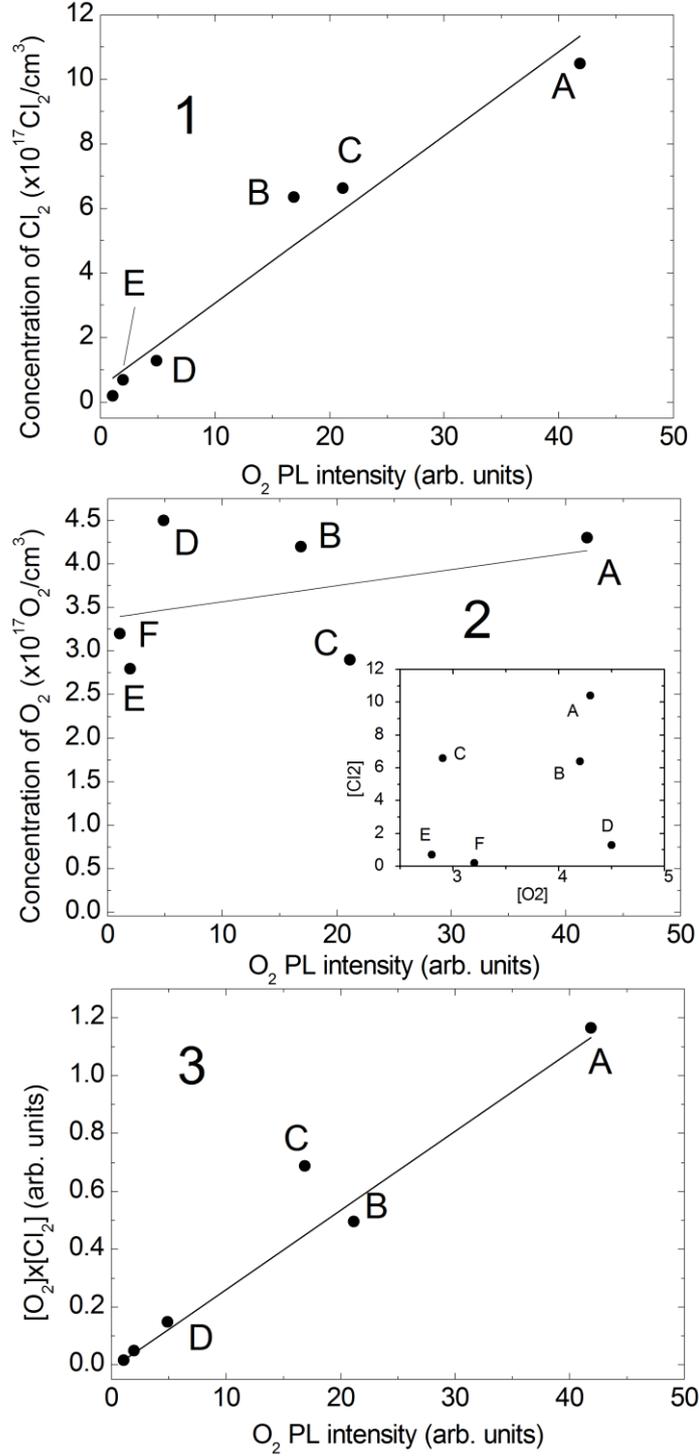

**Figure 7.** Relations between the intensities of UV-excited singlet $O_2$ photoluminescence measured in different samples (A-F, see **Table 1**) and the respective concentrations of interstitial $Cl_2$ (top panel), of interstitial $O_2$ (middle panel), and the product of their concentrations $[O_2]\times[Cl_2]$ (bottom panel). Lines are least-squares linear fits; the outlier data point C in the bottom panel was not included in the fit. The inset in middle panel shows the data of **Table 1**, illustrating the absence of correlation between the concentrations of $O_2$ and $Cl_2$ in the sample set. Excitation: 385 nm (3.22 eV) CW LED.

Decay of the 1272 nm PL intensity was compared using either 385 nm or 766 nm excitation in "high Cl$_2$" sample "A" (**Figure 8**, traces 1,2), and between sample "A" and the "low-Cl$_2$" sample "E", using 766 nm excitation (traces 2 and 3). The kinetics only slightly deviate from exponential ones. To enable comparison between them, the decay constants τ are calculated by linear fit to semi-log plots over the dynamic range of $e^3$ (≈20 times), they are indicated in **Figure 8**.

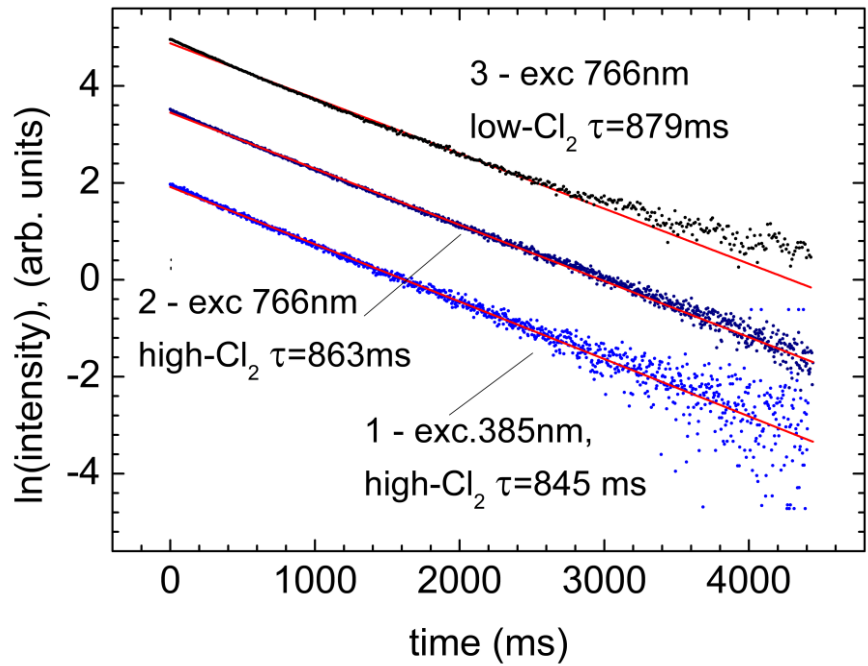

**Figure 8.** Comparison between singlet O$_2$ photoluminescence decay kinetics in cases of energy-transfer excitation (1) and direct excitation (2); and between samples with different concentrations of interstitial Cl$_2$ molecules (traces 2 and 3). The solid lines indicate linear fits.

## 4. Discussion
### 4.1 Optical absorption of dichlorine monoxide(s) and chlorine dioxide

The main effect of F$_2$ laser-irradiation on OA spectrum is a distinct band at 4.7 eV with fwhm 0.83 eV (**Figure 1**). The 4.5 eV-5.5 eV spectral range is very complex in silica, with many OA peaks located there. For example, OA peaks of interstitial ozone (O$_3$) molecules and of oxygen dangling bonds (NBOHC – "non-bridging oxygen hole centers") are both located at 4.8 eV, distinguished only by different fwhm (0.83 eV and 1.07 eV, respectively[21]).

The induced OA peak in **Figure 1** is thus very similar to $O_3$ OA peak, however, it is *red-shifted* by 0.1 eV. As shown by isochronal annealing data (**Figure 2**), the induced OA consists of 2 major components in this spectral region, one of which is stable up to 200°C and another – up to 400°C.

The negative-going "dip" in the difference spectrum (trace 3 in **Figure 1**) at the position of $Cl_2$ OA band, 3.78 eV can be compensated by adding $Cl_2$ OA peak with an amplitude of 0.06 cm$^{-1}$ (traces 4 and 5). Using peak absorption cross section of $Cl_2$ molecules[14] $\sigma=2.6\times10^{-19}$ cm$^2$, it can be estimated that $\approx 2.3\times10^{17}$ $Cl_2$/cm$^3$ have been destroyed by irradiation.

The component of the induced OA, which is destroyed by annealing at 200 °C, has a position and fwhm almost coinciding with those of the ozone band (**Figure 3**). The half-life of gaseous $O_3$ molecules is 1.5 hours at 120 °C and 1.5 s at 250 °C.[22] Interstitial $O_3$ in silica glass has been shown to be destroyed in 10 minutes at 200 °C.[23]. Creation of $O_3$ in excimer laser-irradiated oxygen-rich silica has been observed in many studies. [16,21,23] Therefore, the 4.8 eV OA component in **Figure 3** can be reasonably assigned to $O_3$. Using the $O_3$ peak absorption cross section $\sigma=1.2\times10^{-17}$ cm$^2$ [21], it can be estimated that $2.5\times10^{16}$ $O_3$ molecules were present. This means that less than 10% of the initially present $O_2$ molecules ($4.3\times10^{17}$ cm$^{-3}$) have been converted to $O_3$ by laser irradiation.

While the changes in OA, introduced by the 300 °C annealing step are relatively small (**Figure 2**), noteworthy is the increase of OA in the 3...4 eV range, which indicates a *creation*, instead of destruction, of some absorbing species. At the first glance, the best candidate could be the interstitial $Cl_2$ molecule with its 3.8 eV OA band. However, a closer examination of the corresponding negative peak in difference spectrum (**Figure 3**, trace 3) shows that it cannot be matched by the $Cl_2$ OA band,(3.78 eV/0.70 eV fwhm) and a wider component at lower peak energy ($\approx$3.5 eV) is necessary.

Paramagnetic chlorine dioxide molecule (radical) OClO in gas phase has UV absorption[24], which can be approximated by a Gaussian band at 3.53 eV, fwhm 0.79 eV. This band is drawn in **Figure 3** (trace 4), with amplitude scaled to match the negative dip in the difference spectrum (trace 3). Within the limitations of the relatively poor signal/noise ratio of the experimental data (trace 3) the match can be regarded as satisfactory. Presently there are no OA bands known in silica at this position and with this fwhm. Therefore the weak ~3.5 eV OA band, enhanced by the thermal treatment at 300 °C can be tentatively assigned to creation of interstitial OClO molecules in Cl containing oxygen rich silica. By using the amplitude of the band 0.032 cm$^{-1}$ (trace 4 in **Figure 3**) and peak absorption cross section of OClO,[24] $\sigma=1.2\times10^{-17}$ cm$^2$, their concentration can be estimated as $\sim 3\times 10^{15}$ molecules/cm$^3$.

This assignment is further supported by previous EPR study of Cl-containing $F_2$-laser irradiated silica.[25], where the creation of interstitial OClO (ground state spin S=1/2) molecules was unambiguously proved. OA spectra and their correlation with EPR signal intensities were not studied in that work. However, the isochronal annealing pattern of the EPR signal of OClO (Fig.7 in ref.[25]) is in accord with the behavior of the 3.5 eV OA band, assigned in the present work to OClO: both signals increase significantly on heating between 200 °C and 300 °C, have maximum intensities at 300 °C, and decay on heating to 400 °C (See **Figure 4**, trace 3 below).

The second major induced OA component is destroyed between 300 °C and 400 °C, (**Figure 2**), compared to the interstitial ozone OA band (**Figure 3**, trace 1) it is slightly (0.1 eV) red-shifted and wider (peak at 4.69 eV, fwhm 0.94 eV, **Figure 4**). While this is *per se* only a marginal difference from the ozone OA band in **Figure 3**, there are two additional and more distinct features.

First, Raman spectra show that species giving rise to a line at 955 cm$^{-1}$ are destroyed between 300 °C and 400 °C. (**Figure 5**, the red circle). Resonance Raman peak at 954 cm$^{-1}$ has been assigned to Cl-Cl-O molecules in solid argon [26]; similar Raman peak at 953.5 cm$^{-1}$ has served as a proof of their presence in $F_2$-laser irradiated silica.[7]

Second, the OA spectrum of the component destroyed at 300 °C- 400 °C (**Figure 4**, trace 1) clearly shows a restoration of 3.8 eV OA band due to $Cl_2$ molecules. An assumption that the amplitude of the restored $Cl_2$ OA is equal to that of the irradiation-destroyed absorption (**Figure 1**, trace 4, 0.06 cm$^{-1}$) seems to be reasonable: an addition of an OA band of this magnitude to the difference spectrum (subtracting trace 2 from trace 1 in **Figure 4**) gives just the right amount to eliminate the dip at 3.78 eV (trace 3). This clearly indicates that some Cl-containing species are destroyed and $Cl_2$ molecules are restored at the 400 °C annealing step. In addition, the remaining broad shoulder in the 3 to 4 eV region of trace 3 indicates that the broad OA band at these energies, assigned above to OClO is destroyed at 400 °C.

The thermal destruction of both the OA component 4 in **Figure 4** and the Raman signal of ClClO (**Figure 5**) at 400°C step provides a seemingly strong evidence that interstitial ClClO molecules are responsible for the induced absorption at 4.7 eV and fwhm ≈0.94 eV, and that the laser-assisted destruction of $Cl_2$ molecules proceeds by photolysis of $O_2$ and reaction of O atom with $Cl_2$:

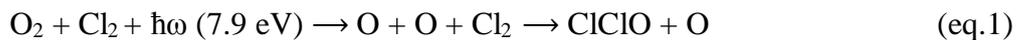

$$O_2 + Cl_2 + \hbar\omega\ (7.9\ eV) \rightarrow O + O + Cl_2 \rightarrow ClClO + O \qquad (eq.1)$$

where the remaining atomic oxygen can form peroxy linkage (Si-O-O-Si bond) in silica network or, alternatively, can enter in some other photolytic reaction.

**Figure 9** shows the comparison between the OA in silica (trace 1) and OA band of ClClO in argon[12] (trace 2) with peak at 4.8 eV. The ~0.1 eV peak shift (4.7 vs. 4.8 eV) can be explained by the effect of the polarizability of the embedding medium (bathochromic shift), which is larger in $SiO_2$ as compared to Ar crystal.

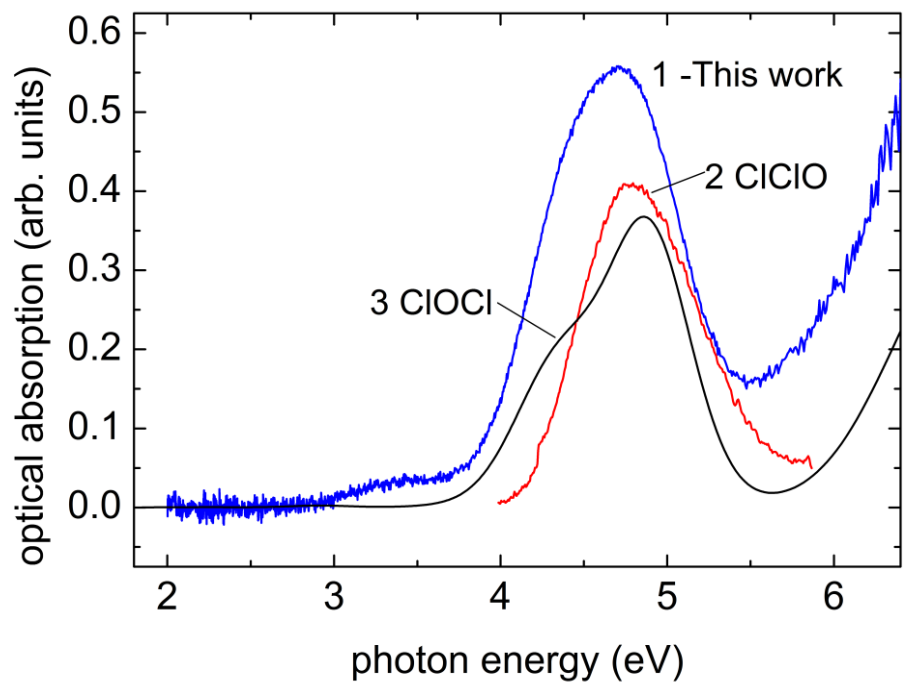

**Figure 9**. Comparison of the optical absorption component, destroyed between 300°C and 400C° (trace 1, the same as trace 3 in Figure 4) with the reported absorption spectra of dichlorine monoxide structural isomers: Cl-Cl-O in Ar matrix (trace 2)[12] and Cl-O-Cl [27] in gas phase (trace 3).

There is, however, a discrepancy in this otherwise consistent picture. The published σ of ClClO OA band ($1.3 \times 10^{-17}$ cm$^2$ [12]) is 50.4 times higher than σ of 3.78 eV $Cl_2$ OA band ($2.58 \times 10^{-19}$ cm$^2$ [14]). Hence the laser-induced destruction of 0.06 cm$^{-1}$ of $Cl_2$ OA band by converting $Cl_2$ to ClClO should introduce a ClClO band with an amplitude of 3.0 cm$^{-1}$, ca. 5× larger than the actually created OA band (**Figure 9**, trace 1). There exists a structural isomer of ClClO, the symmetric ClOCl molecule, which is thermodynamically more stable and has lower σ, $1.94 \times 10^{-18}$ cm$^2$ for OA peak at 4.86 eV, [27] only 4 times larger than that of $Cl_2$. The shape of ClOCl OA is depicted by trace 3 in **Figure 9**. If all destroyed $Cl_2$ molecules would form ClOCl, then the amplitude of the induced OA band would be only ≈0.24 cm$^{-1}$, much smaller than the amplitude (0.55 cm$^{-1}$) of the actually induced OA band. Therefore we suggest that the induced OA band in **Figure 9** originates from a mixture of both structural isomers, Cl-Cl-O and Cl-O-Cl. Only one of them, ClClO is detected in Raman spectra (**Figure 5**), because the Raman band of Cl-O-Cl is weaker and lays in the "non-flat" spectral region (bands at 634 and 672 cm$^{-1}$ [26], where an accurate measurement of low-intensity difference spectra is unreliable. The difference in spectral shapes of traces 1 and a linear combination of traces 2 and 3 (**Figure 9**) can be attributed to bathochromic shifts and inhomogeneous broadening (typically of ~0.1eV) in silica glass matrix.

**4.2 Cl$_2$-O$_2$ energy transfer**

The luminescence of singlet state O$_2$ molecules ("$^1$O$_2$") is notoriously difficult to photo-excite directly, because the transitions from the ground state to its 2 lowest electronic excited states are strongly forbidden, both by spin and by parity rules[18], and the reasonably strongly allowed transitions ($\sigma > 10^{-20}$ cm$^2$) start only at photon energies in the vacuum UV range ($\hbar\omega > 6.5$ eV).[20] Therefore, the prevailing majority of the numerous $^1$O$_2$ studies employs activators ("photosensitizers"), e.g., metalorganic molecules, to transfer the absorbed photo-excitation energy to the nearby O$_2$ molecule. Studies of direct excitation to the two lowest electronic excited states (optical transitions at 0.97 eV and 1.62 eV) are relatively scarce. In the case of interstitial O$_2$ in silica the situation is reverse: in numerous past studies of bulk glass (see paper [19] for references) or nanoparticles,[28] only the direct photoexcitation of O$_2$ has been reported for photon energies below 6.4 eV. However, $^1$O$_2$ PL in silica is excited also in processes involving excitons or electron-hole recombination, caused by electron-[29] or X-ray[30] irradiation, and in photolysis of interstitial O$_3$ molecules.[21] Therefore, the observation of PL band of $^1$O$_2$ under 355 nm (3.49 eV) pulsed Nd-YAG laser excitation (**Figure 6**, trace 1), which falls into photon energy range, void of O$_2$ excited states[20], led one to suspect 2-photon excitation. However, the sublinear excitation power dependence (Section 3.3) and the excitation of $^1$O$_2$ PL by CW light at 385 nm (**Figure 6**, trace 2) clearly indicates that an energy transfer to O$_2$ molecule from some "activator" species absorbing in the 3-4 eV range takes place. The low-energy wing of the OA band of interstitial Cl$_2$ molecules (**Figure 1**, trace 1) covers this range. Multiple different silica glass samples, apart from those listed in **Table 1,** having different stoichiometries and irradiation histories were tested, and it was found that the most intense UV-excited PL of $^1O_2$ is observed in samples containing both $O_2$ and $Cl_2$. That strongly hints that energy transfer from $Cl_2$ to $O_2$ takes place.

Absorption of photon of 3...4 eV energy by $Cl_2$ molecule transfers it to $^1\Pi_u$ excited state, which is repulsive,[31] however, the 3.8 eV OA band is not photobleached as both Cl atoms are held together due to the "cage effect" of silica glass.[7] The lower-lying $^3\Pi_u$ triplet state of $Cl_2$ has a shallow minimum[31] and at low temperatures gives rise to a structured "comb-shaped" near-infrared luminescence of $Cl_2$ in silica.[7] Its non-structured part can be approximated by a Gaussian with peak at 1.215 eV and fwhm 0.39 eV[8], which means that $Cl_2$ PL spectrum has a significant intensity (35% from the peak value) at the wavelength of the strongest [7,18] NIR absorption of the $O_2$ molecule, ~1270 nm (0.976 eV), corresponding to the excitation to the $^1\Delta_g$ singlet state. In addition, 5% of the $Cl_2$ PL intensity overlaps with the next higher-energy $O_2$ absorption band at 765 nm (1.62 eV), providing a possibility of energy transfer to another singlet ($^1\Sigma_g^+$) state of $O_2$. Therefore, the lowest triplet ($^3\Pi_u$) excited state of $Cl_2$ is a reasonable candidate for the energy-donor state. The energy transfer to the $O_2$ molecule is most probably described as

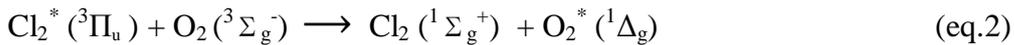

$$Cl_2^* \,(^3\Pi_u) + O_2\,(^3\Sigma_g^-) \longrightarrow Cl_2\,(^1\Sigma_g^+) + O_2^*\,(^1\Delta_g) \qquad (eq.2)$$

where the remaining excess energy of excited $Cl_2$ is absorbed by vibrations of $Cl_2$ or $O_2$.

Two basic energy transfer mechanisms are generally considered: the dipole-dipole (Förster) mechanism and the direct electron exchange (Dexter) mechanism.[32] The Förster mechanism works over larger distances up to r = 10 nm, the energy transfer rate decreases with distance as $\sim 1/r^6$. This energy-transfer mechanism is used in organic macromolecules to measure the distance between the energy donor and acceptor sites[33]. However, it is efficient only for dipole-allowed electronic transitions, but the transition to the excited singlet state of $O_2$ from its ground state is strictly forbidden. Therefore the excitation of ground-state $O_2$ to singlet state must occur by Dexter mechanism,[32] which has much shorter transfer range and requires some overlap of the energy donor and acceptor wavefunctions.

The process of energy transfer from energy donor D to ground-state $O_2$ molecule must obey the spin conservation rule: the total spin of the system (D+$O_2$) must remain unchanged by the energy-transfer process. This is possible by the "triplet annihilation" mechanism, where triplet state excited sensitizer transfers its energy selectively to those ground-state triplet $O_2$ molecules, which have opposite orientations of their S=1 total spin. Other spin-allowed mechanism ("triplet exchange") is also known, which involves a partial energy transfer from the excited singlet state of the donor to triplet $O_2$, and results in a lower-energy excited triplet state donor and excited singlet $O_2$. However this mechanism requires relatively large (at least 1 eV) singlet-triplet splitting in the donor species and is relatively rare; the "triplet annihilation" is by far the most common singlet $O_2$ activation mechanism.[34] The energy transfer to $O_2$ from the triplet state of $Cl_2$ most likely occurs by this mechanism, as suggested by **eq.2**. Hence it follows that $O_2$ and $Cl_2$ molecules giving activated UV-excited PL of $^1O_2$ are direct or next-closest neighbors.

### 4.3 Distribution of $O_2$ and $Cl_2$ in silica network

**Figure 7** (panel 1) shows that there is a reasonably strong correlation between the UV-excited $^1O_2$ PL intensity and concentration of $Cl_2$ in different samples, where $Cl_2$ concentration varies over 10 times. In contrast, the correlation with the $O_2$ content is less clear (panel 2). However, this discrepancy may simply reflect the circumstance that the range of the available $O_2$ concentrations was much smaller (change of only 1.6 times) compared to the range of $Cl_2$ concentrations . The best, almost linear correlation, apart from 1 outlier point, is found between the singlet $O_2$ PL intensity and the product of $Cl_2$ and $O_2$ concentrations (panel 3). This finding is compatible with a random distribution of $Cl_2$ and $O_2$ in glass matrix, since at low concentrations ($10^{17}..10^{18}$ molecules/cm$^3$) the probability of forming close pair is proportional to the product of concentrations. A larger range of $O_2$ and $Cl_2$ concentrations should be studied in order to determine more exactly if the mutual distribution of $Cl_2$ and $O_2$ interstitial molecules is indeed purely random.

The number of large (diameter> 0.5nm) interstices, capable of accommodating $O_2$ or $Cl_2$ molecules in silica glass can be roughly estimated as $\approx 10^{22}/cm^3$, using the data of molecular dynamics studies,[35] which give the total volume, occupied by such voids as 32%. With $10^{18}$ molecules/$cm^3$ distributed in these voids, the probability of having 2 neighboring voids occupied is >$10^{-4}$. This number (~$10^{14}$ – $10^{15}$ pairs/$cm^3$) is large enough to be detected by luminescence and thus is consistent with the current observations. On the other hand, the OA data (**Figure 1**) show that a much larger number, of $Cl_2$ molecules ($2\times10^{17}$ $cm^{-3}$) react with oxygen atoms in $F_2$ laser –induced reaction. This is possible, either if $O_2$ and $Cl_2$ are paired (e.g., in very large interstitial voids), contrary to the random-distribution discussed above, or if the laser photolysis-generated O atoms are mobile enough to travel longer paths to distant $Cl_2$ molecules. It is generally agreed[36] that thermalized O atom diffuses by exchange with network O atoms, and forms a peroxy linkage Si-O-O-Si. This diffusion is very slow at the room temperature. However, it has been demonstrated with isotopically $^{18}O_2$ - enriched samples that part of photolysis-generated $^{18}O$ atoms while traveling long distances to form $O_3$ molecules, do not undergo exchange with network oxygens.[23] By analogy, similar process can be involved in $Cl_2O$ formation.

The possibility that both $Cl_2$ and $O_2$ molecules could initially reside in a single large void, can be additionally examined by PL decay measurements (**Figure 8**). The decay of singlet $O_2$ PL in "dry" silica (decay constant $\tau \approx$ 850 ms) at room temperature is the slowest among observations of $O_2$ PL in any condensed matter matrix,[19] because the rate of $^1O_2$ non-radiative relaxation in $SiO_2$ matrix is record-low. All other matrices cause higher non-radiative rates. Therefore, an additional perturbation by a closely contacting activator is *generally* expected to shorten the decay kinetics. However, **Figure 8** demonstrates that $^1O_2$ PL kinetics is practically the same in the cases of energy-transfer excitation (385 nm, trace 1) and direct photoexcitation of $O_2$ (766 nm, trace 2). This finding is supportive of $Cl_2$ and $O_2$ having little interaction and is in accord with the assumption that both molecules are located in separate voids in silica glass network. However, it must be kept in mind that $Cl_2$ can be a relatively low efficiency quencher of $^1O_2$ PL in silica compared to, e.g., silanol group $\equiv$Si−O−H, and therefore the presence of closely located $Cl_2$ does not necessarily mean strong quenching. The quenching of $^1O_2$ PL is often dominated by electronic-to-vibrational energy transfer, which in case of $Cl_2$ may have low rate, because the vibrational frequency of $Cl_2$ in silica (546 cm$^{-1}$ [7,8]) is significantly lower than that of silanol ($\approx$3700 cm$^{-1}$). A previous study [37] has shown that chlorine impurities in the form of $\equiv$Si−Cl groups (up to $5\times10^{19}$ cm$^{-3}$) have no measurable effect on quenching of $^1O_2$ PL in silica.

## 5. Conclusion

The main result of this study is the identification of radiation-induced optical absorption band in the 4.7 eV region due to the formation of dichlorine monoxide $Cl_2O$ molecules in reactions molecular chlorine with oxygen atoms. While only one of the two structural isomers of dichlorine monoxide, a ClClO molecule was identified by Raman spectrum, the contribution of the other isomer, ClOCl to the 4.7 eV optical absorption band can be significant. These data may help to decipher the often very complex optical absorption spectrum of irradiated silica glass in the "difficult" 4.5-5.5 eV spectral range, where a number of closely spaced OA bands due to intrinsic defects (NBOHC, peroxy radical, silicon oxygen deficiency centers SiODC(II), interstitial ozone) overlap.

An additional finding is the weak optical absorption band centered at 3.5 eV with fwhm 0.8 eV. Its position and width are similar to those of chlorine dioxide, OClO molecule. Its tentative assignment to interstitial OClO in silica is additionally supported by similar intensity changes of this optical absorption band and of the previously published EPR signal of OClO[25] during isochronal annealing series. This band can contribute to the non-structured absorption losses in irradiated optical fibers, generally increasing with photon energy in blue and UV regions. It is well-established that these losses in optical fibers are related to chlorine impurities[10]; the idea that chlorine dioxide could be the responsible species has been proposed long ago[38]. Our present data support this proposal.

Chlorine impurities are nearly omnipresent in all silica glasses produced from $SiCl_4$ or dried by $Cl_2$ in the porous "soot" preform phase. In contrast, excess $O_2$, necessary for the formation of $Cl_2O$ is normally not present in commercial glasses. However, it is a typical product of radiolysis of oxide glasses (e.g.,[39]). Excess $O_2$ may be added deliberately, because it is found to increase the radiation toughness of optical fibers in the NIR spectral region by suppressing the generation of self-trapped holes ($STH_1$).[40] And most commonly, it is inadvertently loaded in sub-surface layer of silica by high-temperature processing under atmospheric conditions.[41] The defect generation in the subsurface layer of silica optics is a crucial factor in high-power laser applications.

The present data are consistent with the model of random distribution of $Cl_2$ and $O_2$ molecules in silica glass network and of absence of both pair formation and of significant clustering of these molecules in large voids. However, a study over wider ranges of $O_2$ and $Cl_2$ concentrations would be useful to finally verify this model. The energy transfer between $Cl_2$ and $O_2$ found in the present study may serve as an additional tool to indicate the proximity of both molecules.

**Acknowledgements**

The support from Latvian Science Council project lzp-2018/1-0289 is acknowledged. K.K. was partially supported by the Collaborative Research Project of Laboratory for Materials and Structures, Tokyo Institute of Technology.